\documentclass[prl, twocolumn, floatfix, superscriptaddress, longbibliography, preprintnumbers]{revtex4-1}

\usepackage{amsmath, amssymb}
\usepackage{graphicx}
\usepackage{color}

\newcommand{\ket}[1]{{| #1 \rangle}}
\newcommand{\bra}[1]{{\langle #1 |}}
\newcommand{\ee}{{\rm e}}
\newcommand{\ii}{{\it i}}
\newcommand{\dd}{{\rm d}}

\newcommand{\sinc}{{\rm sinc}}

\newcommand{\cm}{{\rm cm}}
\newcommand{\rel}{{\rm rel}}

\newcommand{\SP}{{\rm RI}}
\newcommand{\Le}{{\rm L}}
\newcommand{\Ri}{{\rm R}}
\newcommand{\Early}{{\rm e}}
\newcommand{\Late}{{\rm l}}
\newcommand{\deloc}{\Delta x_{\rm rel}}

\newcommand{\bs}{\boldsymbol} 

\begin{document}

\title{Disorder-robust entanglement transport}

\author{Clemens Gneiting}
\email{clemens.gneiting@riken.jp}
\affiliation{Theoretical Quantum Physics Laboratory, RIKEN Cluster for Pioneering Research, Wako-shi, Saitama 351-0198, Japan}
\author{Daniel Leykam}
\affiliation{Center for Theoretical Physics of Complex Systems, Institute for Basic Science (IBS), Daejeon 34126, Republic of Korea}
\author{Franco Nori}
\affiliation{Theoretical Quantum Physics Laboratory, RIKEN Cluster for Pioneering Research, Wako-shi, Saitama 351-0198, Japan}
\affiliation{Department of Physics, University of Michigan, Ann Arbor, Michigan 48109-1040, USA}

\date{\today}

\begin{abstract}
We study the disorder-perturbed transport of two noninteracting entangled particles in the absence of backscattering. This situation is, for instance, realized along edges of topological insulators. We find profoundly different responses to disorder-induced dephasing for the center-of-mass and relative coordinates: While a mirror symmetry protects even highly delocalized relative states when resonant with the symmetry condition, delocalizations in the center of mass (e.g. two-particle ($N=2$) $N00N$ states) remain fully sensitive to disorder. We demonstrate the relevance of these differences to the example of interferometric entanglement detection. Our platform-independent analysis is based on the treatment of disorder-averaged quantum systems with quantum master equations.
\end{abstract}

\preprint{\textsf{published in Phys.~Rev.~Lett.~{\bf 122}, 066601 (2019)}}

\maketitle

\section{Introduction}

Uncontrolled perturbances (disorder) can significantly modify the expected or, for that matter, desired transport behavior of quantum particles. This does not only hold for their overall mobility properties, which have traditionally been intensively investigated \cite{Lifshits1988introduction, Rammer1991quantum, Beenakker1997random}, but also for the detailed phase information encoded in quantum states. The latter, in turn, controls the particles' ability to interfere and thus underlies their utilization in quantum experiments and technologies.

The preservation of phase relations during transport is a delicate task, even if backscattering, localization, and environmental decoherence are negligible. In the case of single particles, it has been shown that disorder-induced dephasing can, depending on state specifications and dispersion, significantly reduce the fidelity of interference applications, possibly putting their successful deployment at stake \cite{Gneiting2017quantum, Gneiting2017disorder}.

Several quantum aspects, such as entanglement and particle statistics, only arise for two or more particles, causing genuine quantum behavior, such as nonclassical correlations, quantum teleportation, (anti-)bunching, etc.~\cite{Lahini2010quantum, Peruzzo2010quantum,Crespi2013anderson, Giuseppe2013einstein, Poulios20142D,Nicola2014quantum,Solntsev2014biphoton,Wang2018metasurface}. Again, phase information plays here a crucial role, and analyzing the effect of disorder beyond localization is important for potential applications. On the other hand, new insights into the interplay between the impact of disorder, entanglement, and particle statistics are expected to emerge.

\begin{figure}[htb]
	\includegraphics[width=0.99\columnwidth]{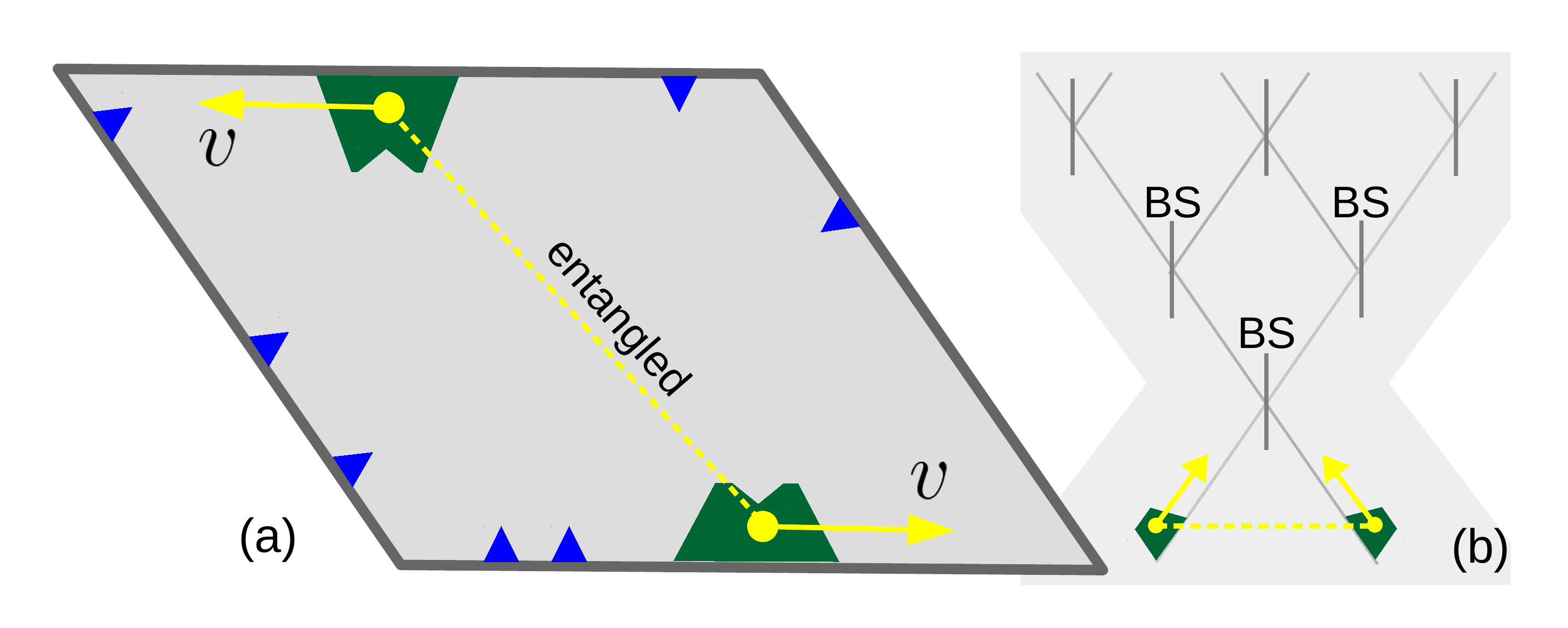}
	\caption{\label{Fig:entangled_edge-mode_pair} (a) Disorder-perturbed transport of two-particle entangled edge modes. While topologically protected against backscattering, perturbations (schematic, blue) along the paths of the particles still cause disorder-induced dephasing, deteriorating the possibility to detect and/or harness their entanglement. (b) Disorder-induced dephasing degrades, e.g., two-particle coherence effects, such as (anti-)bunching at beam splitters. If cascaded, the disorder impact accumulates.}
\end{figure}

In this article, we systematically study the effect of disorder potentials on the backscattering-free transport of two-particle entangled states, cf.~Fig.~\ref{Fig:entangled_edge-mode_pair}, relevant to topological edge modes in photonic and condensed matter systems~\cite{topological_photonics_review,topological_insulator_review}. Our analytical treatment of the disorder impact in terms of ensemble-averaged quantum states reveals a mirror symmetry in the response to disorder, which can be exploited to achieve disorder-robust transport of entangled states. We stress that this robustness lies in the phase information of the two-particle state and emerges when both particles simultaneously reside in the same pairs of spatial locations; it cannot be understood simply in terms of the absence of backscattering of single particle or $N00N$ states~\cite{Mittal2016topologically,Rechtsman2016topological,Mittal2017topologically}. Our findings, along with a similar effect in the response of two identical particles to environmental dephasing \cite{Perez2018endurance}, thus demonstrate potential to enhancing topological protection using multiparticle states.

\section{Disorder-averaged evolution}

We consider backscattering-free propagation of two spinless quantum particles in one dimension, described by a common, constant drift velocity $v$. This generalizes the single-particle case discussed in Ref.~\cite{Gneiting2017disorder}. To be general, we consider two distinguishable particles. This encompasses identical particles, either by appropriately symmetrizing initial states, or if additional internal degrees of freedom, in the case of photons, e.g., polarization, lift the symmetry constraints on the spatial state component.

The Hamiltonian in the presence of a disorder potential then reads ($v>0$)
\begin{align} \label{Eq:Disorder-perturbed_Hamiltonian}
\hat{H}_{\varepsilon} = v (\hat{p}_1+\hat{p}_2) + V_{\varepsilon}(\hat{x}_1) + V_{\varepsilon}(\hat{x}_2) ,
\end{align}
where $x$ describes the position along the edge. The (multi-)index $\varepsilon$ labels different disorder realizations, which may occur with probability $p_{\varepsilon}$ (for simplicity we write integrals throughout, e.g., $\int \dd \varepsilon \, p_{\varepsilon} = 1$).

Both particles encounter the same, homogeneous, disorder potential $V_{\varepsilon}(\hat{x}) = \int_{-\infty}^{\infty} \dd x \, V_{\varepsilon}(x) \ket{x}\bra{x}$, characterized by translation-invariant two-point correlations $C(x-x') \equiv \int \dd \varepsilon \, p_{\varepsilon} \, V_{\varepsilon}(x) V_{\varepsilon}(x') = \int_{-\infty}^{\infty} \dd q \, \ee^{\frac{\ii}{\hbar} q (x-x')} G(q)$,
where the distribution $G(q)$ (see also Refs.~\cite{Gneiting2016incoherent, Gneiting2017quantum}) describes the correlations in momentum space. For simplicity, the disorder potential may also vanish on average, $\int \dd \varepsilon \, p_{\varepsilon} \, V_{\varepsilon}(x) = 0$, such that the average Hamiltonian reads $\hat{\overline{H}} \equiv \int \dd \varepsilon \, p_{\varepsilon} \, \hat{H}_{\varepsilon} = v (\hat{p}_1+\hat{p}_2)$.

In the limit of weak disorder, the dynamics of the disorder-averaged state $\overline{\rho}(t) = \int \dd \varepsilon \, p_{\varepsilon} \ee^{-\frac{\ii}{\hbar} \hat{H}_{\varepsilon} t} \rho_0 \ee^{\frac{\ii}{\hbar} \hat{H}_{\varepsilon} t}$ can be described by a quantum master equation \cite{Gneiting2016incoherent, Kropf2016effective, Gneiting2017quantum, Gneiting2017disorder, Gneiting2018lifetime}, which is perturbative to second order in the disorder potential \cite{Gneiting2017quantum}. Abbreviating $\mathcal{L}(\hat{L}, \rho) \equiv \hat{L} \rho \hat{L}^{\dagger} - \frac{1}{2} \hat{L}^{\dagger} \hat{L} \rho - \frac{1}{2} \rho \hat{L}^{\dagger} \hat{L}$, and using $C(x-x')$ and $\hat{\overline{H}}$, we obtain the disorder-dressed evolution equation
\begin{align} \label{Eq:perturbative_master_equation}
\partial_t \overline{\rho}(t) =& -\frac{\ii}{\hbar} [\hat{\overline{H}}, \overline{\rho}(t)] \\
& +\sum_{\alpha \in \{ \pm 1 \}} \frac{2 \alpha}{\hbar^2} \int_{-\infty}^{\infty} \dd q \, G(q) \int_{0}^{t} \dd t' \mathcal{L}\big(\hat{L}_{q, t'}^{(\alpha)}, \overline{\rho}(t)\big) , \nonumber
\end{align}
with the Lindblad operators $\hat{L}_{q, t}^{(\alpha)} = \frac{1}{2} \big[ \hat{V}_{q} + \alpha \, \hat{\tilde{V}}_{q}(t) \big]$, where $\hat{V}_q = \ee^{\frac{\ii}{\hbar} q \hat{x}_1} + \ee^{\frac{\ii}{\hbar} q \hat{x}_2}$ and $\hat{\tilde{V}}_q(t) = \ee^{-\frac{i}{\hbar} v q t} \hat{V}_q$. Note that the $\hat{V}_q$ describe simultaneous, coherent momentum kicks of both particles. This follows from the fact that both particles encounter the same disorder potential, introducing correlations relevant at common locations of the two particles.

Recasting Equation~(\ref{Eq:perturbative_master_equation}) in terms of center-of-mass, $x_\cm = (x_1+x_2)/2$, and relative coordinate, $x_\rel = x_1-x_2$, yields
\begin{align} \label{Eq:master_equation_evaluated}
\partial_t \overline{\rho}(t) =& -\frac{\ii}{\hbar} [v \hat{p}_\cm, \overline{\rho}(t)] \\
&+ \int_{-\infty}^{\infty} \!\!\! \dd q \; \frac{8 t G(q)}{\hbar^2} \sinc \! \left[ \frac{q v t}{\hbar} \right] \mathcal{L}\big(\hat{L}_q, \overline{\rho}(t)\big) , \nonumber
\end{align}
with the Lindblad operators $\hat{L}_q = \ee^{\frac{i}{\hbar} q \hat{x}_\cm} \cos \left[ \frac{q \hat{x}_\rel}{2 \hbar} \right]$.
We find that center-of-mass and relative coordinate are affected differently by the disorder potential: While the former behaves similarly to a disorder-pertubed single-particle edge state (cf.~Ref.~\cite{Gneiting2017disorder}), the latter experiences coherent momentum kicks in opposing directions. The solution of Eq.~(\ref{Eq:master_equation_evaluated}) reads [$G(-q) = G(q)$] $\bra{x_\cm, x_\rel} \overline{\rho}(t) \ket{x_\cm',x_\rel'} =$
\begin{subequations} \label{Eq:master_equation_solution}
	\begin{align}
	\bra{x_\cm-v t, x_\rel} \rho_0 &\ket{x_\cm'-v t, x_\rel'} \\
	&\times \exp \left[ -\overline{F}_t(x_\cm, x_\rel, x_\cm', x_\rel') \right] , \nonumber
	\end{align}
where $\rho_0$ describes an arbitrary initial state, and with the disorder influence
	\begin{align} \label{Eq:disorder_influence}
	\overline{F}_t(x_\cm, x_\rel, x_\cm', x_\rel') =& \frac{4 t^2}{\hbar^2} \int \dd q \, G(q) \, \sinc^2 \left[ \frac{q v t}{2 \hbar} \right] \nonumber \\
	\times \Bigg\{ \frac{1}{2} \cos^2 \left[ \frac{q x_\rel}{2 \hbar} \right] &+ \frac{1}{2} \cos^2 \left[ \frac{q x_\rel'}{2 \hbar} \right] \\
	- \cos \left[ \frac{q (x_\cm-x_\cm')}{\hbar} \right]& \cos \left[ \frac{q x_\rel}{2 \hbar} \right] \cos \left[ \frac{q x_\rel'}{2 \hbar} \right] \Bigg\} . \nonumber
	\end{align}
\end{subequations}

\begin{figure}[htb]
	\includegraphics[width=0.99\columnwidth]{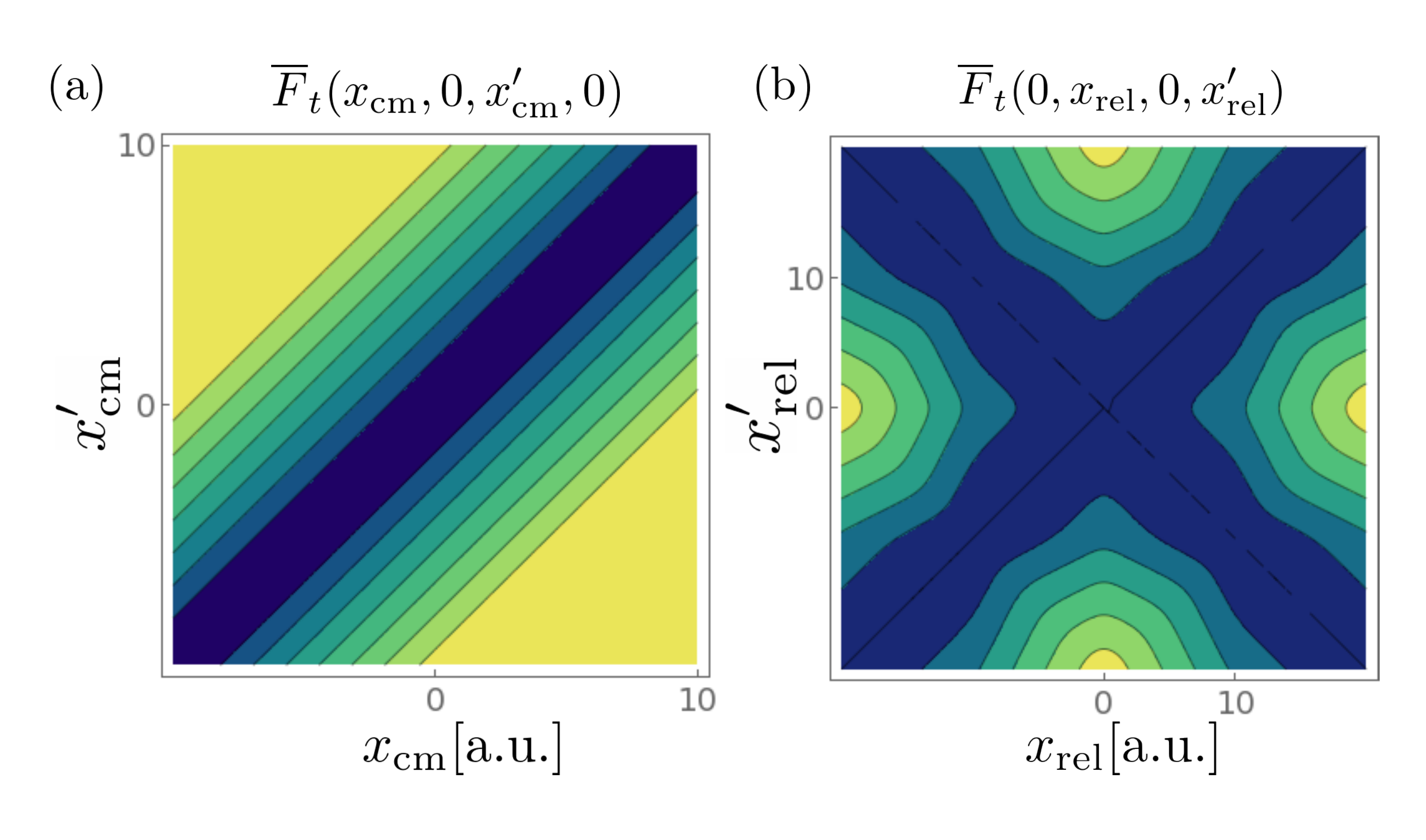}
	\caption{\label{Fig:mirror_point_robustness} Disorder influence (\ref{Eq:disorder_influence}) for a pair of propagating edge modes [Gaussian disorder correlations, $t=10\ell/v$, values increase from $0$ (blue)]. (a) While the center-of-mass coordinate shows a dephasing behavior similar to a single particle, with coherences between coordinates $x_{\cm}$ and $x_{\cm}'$ homogeneously degrading with their increasing separation, (b) coherences between mirror points $x_{\rel}$ and $-x_{\rel}$ of the relative coordinate remain unaffected by disorder-induced dephasing, regardless of their spatial separation. This is because both particles acquire the same disorder phases which thus cancel out.}
\end{figure}

Note that Eq.~(\ref{Eq:disorder_influence}) reduces to the single-particle case when evaluated for $x_\rel=x_\rel'=0$, describing a decoherence cone, with coherences between remote points $x_\cm$ and $x_\cm'$ decaying homogeneously with increasing spatial separation, cf.~\cite{Gneiting2017disorder}. In the relative coordinate, however, one finds, for $x_\cm=x_\cm'$, that coherences of mirror points $x_\rel$ and $-x_\rel$ are robust against disorder effects, independently of their spatial separation, see Fig.~\ref{Fig:mirror_point_robustness}. This is because, in this instance, both particles simultaneously reside in the same pair of spatial locations, such that the different phases acquired from the disorder potential cancel each other exactly (or rather cause an irrelevant global phase). This insight will guide us to identify spatially delocalized disorder-robust entangled states. We remark that this symmetry can be related to the permutational invariance of the Hamiltonian~(\ref{Eq:Disorder-perturbed_Hamiltonian}).

While Solution (\ref{Eq:master_equation_solution}) holds for arbitrary correlations $C(x)$, we can evaluate the disorder influence for generic Gaussian correlations $C(x) = C_0 \, \exp \left[- \left(\frac{x}{\ell} \right)^2 \right]$, where $\ell$ denotes the correlation length. With $G(q) \equiv \frac{1}{2 \pi \hbar} \int_{-\infty}^{\infty} \dd x \, \ee^{-\frac{\ii}{\hbar} q x} C(x) = \frac{C_0 \ell}{2 \sqrt{\pi} \hbar} \ee^{-\frac{1}{4}\left( \frac{q \ell}{\hbar} \right)^2}$, one then obtains $\overline{F}_t(x_\cm, x_\rel, x_\cm', x_\rel') = \sum_{\sigma_1,\sigma_2 =\pm1} \overline{F}_t^{(1)}([x_\cm-x_\cm']+\sigma_1 [x_\rel+\sigma_2 x_\rel'])$, with the single-particle disorder influence $\overline{F}_t^{(1)}(x) = \frac{C_0 \ell^2}{\hbar^2 v^2} \left\{ 2 \overline{f} \left( \frac{v t}{\ell} \right) + 2 \overline{f} \left( \frac{x}{\ell} \right) - \overline{f} \left( \frac{x - v t}{\ell} \right) - \overline{f} \left( \frac{x + v t}{\ell} \right) - 2 \overline{f}(0) \right\}$ and $\overline{f}(x) = x \, {\rm erf}(x) + (\ee^{-x^2}/\sqrt{\pi})$ \cite{Gneiting2017disorder}. Hereafter, we always assume Gaussian correlations when the disorder influence is evaluated.

\section{Two-particle interference}

To assess the disorder robustness at mirror points, we now investigate how the disorder-perturbed edge propagation affects entangled states supporting two-particle interference. To this end, we consider superposition states delocalized in the relative coordinate,
\begin{align} \label{Eq:Modular_momentum_entangled_state}
	&\ket{\Psi_{\SP}} = \sqrt{\frac{\sigma_{p,\cm}}{\pi \hbar \sigma_{x,\rel}}} \int \dd x_\cm \, \dd x_\rel \, \ket{x_\cm} \otimes \ket{x_\rel} \\
	&\times \ee^{-\frac{\sigma_{p,\cm}^2 x_\cm^2}{\hbar^2}} \frac{1}{\sqrt{2}} \left( \ee^{-\frac{(x_\rel-x_{\Le})^2}{4 \sigma_{x,\rel}^2}} + e^{i \varphi} \, \ee^{-\frac{(x_\rel-x_{\Ri})^2}{4 \sigma_{x,\rel}^2}} \right) , \nonumber
\end{align}
where we assume that the spatial delocalization $\deloc \equiv |x_{\Le}-x_{\Ri}|$ of the two state components well exceeds their uncertainty, $\deloc \gg \sigma_{x,\rel}$. The mirror condition is fulfilled if $x_{\Le}=-x_{\Ri}$. Note that it must be met by identical particles, unless additional degrees of freedom lift the symmetry constraints. The phase $\varphi$ may accommodate for (anti-)symmetric states under particle exchange. For simplicity, we assume $\varphi=0$.

The bipartite entangled state (\ref{Eq:Modular_momentum_entangled_state}) supports two-particle interference in the relative momentum, as seen by inspection of its momentum distribution, $P_\SP(p_\cm,p_\rel) \equiv |\bra{p_\cm,p_\rel} \Psi_{\SP}\rangle|^2 \propto$
\begin{align} \label{Eq:Nonlocal_interference_pattern}
\ee^{-\frac{p_\cm^2}{2 \sigma_{p,\cm}^2}} \ee^{-\frac{2 \sigma_{x,\rel}^2 p_\rel^2}{\hbar^2}} \left\{ 1 + \cos \left[ \frac{p_\rel \deloc}{\hbar} \right] \right\} .
\end{align}
In this sense, it generalizes Young interference experiments to the bipartite case \cite{Gneiting2013nonlocal}. Such interference pattern could, for instance, be measured by guiding the state into a Mach-Zehnder interferometer arrangement.

The interference pattern (\ref{Eq:Nonlocal_interference_pattern}), characterized by the delocalization $\deloc$, occurs irrespectively of whether the mirror condition is met or not. Moreover, it is unaffected by additional correlations within the two superposed state components, which could be replaced by separable states. In that sense, we can, if the mirror condition is met, associate (\ref{Eq:Modular_momentum_entangled_state}) with the time-bin entangled state $\frac{1}{\sqrt{2}}(\ket{\Early}_1 \ket{\Late}_2 + e^{i \varphi} \ket{\Late}_1 \ket{\Early}_2)$ \cite{Brendel1999pulsed, Gneiting2008bell, Gneiting2009nonclassical, Mittal2016topologically, Mittal2017correlations}, where $\ket{\Early}$ and $\ket{\Late}$ denote ahead-moving (``early'') and following (``late'') wave packets.

To detect the entanglement of Eq.~(\ref{Eq:Modular_momentum_entangled_state}), we employ an interferometric entanglement criterion, which is formulated in terms of the modular variables $\overline{x} = x \mod \deloc$ and $\overline{p} = (p+h/2 \deloc) \mod (h/\deloc) - h/2 \deloc$, and their respective integer components $N_x = [(x-v t)-\overline{x}]/\Delta x_\rel$ (using a comoving origin of the coordinate system) and $N_p = (p-\overline{p})\Delta x_\rel / h$ \cite{Aharonov1969modular, Gneiting2011detecting}. With $N_{x, {\rm tot}} \equiv N_{x,1}+N_{x,2}$ and $\overline{p}_\rel \equiv \overline{p}_1-\overline{p}_2$, the entanglement criterion reads \cite{Gneiting2011detecting, Gneiting2013nonlocal} $\langle (\Delta \hat{N}_{x, {\rm tot}})^2 \rangle + \frac{\deloc^2}{h^2} \langle (\Delta \hat{\overline{p}}_\rel)^2 \rangle < 2 C_{\hat{N}_x,\hat{\overline{p}}}$, where the constant $C_{\hat{N}_x,\hat{\overline{p}}}$ is obtained numerically to $C_{\hat{N}_x,\hat{\overline{p}}} \approx 0.078$. A state which satisfies the criterion is certified to be entangled. We note that applying this entanglement criterion presupposes distinguishable particles, which we assume now for demonstrational purposes. The interference is also present for identical particles.

For the unperturbed superposition state (\ref{Eq:Modular_momentum_entangled_state}), we have ($\deloc \gg \sigma_{x,\rel}, \hbar/\sigma_{p,\cm}$) $\langle (\Delta \hat{N}_{x, {\rm tot}})^2 \rangle \approx 0$ and $\frac{\deloc^2}{h^2} \langle (\Delta \hat{\overline{p}}_\rel)^2 \rangle = \frac{1}{6} [1-S_2(2)]$, with $S_2(2) = \frac{3}{\pi^2} \approx 0.304$. The left-hand side thus evaluates as 0.117, which is well below the threshold value of 0.156, classifying the state as entangled. An Einstein-Podolsky-Rosen entangled state, on the other hand, corresponding to a single superposition branch in (\ref{Eq:Modular_momentum_entangled_state}), would yield $\frac{\deloc^2}{h^2} \langle (\Delta \hat{\overline{p}}_\rel)^2 \rangle = \frac{1}{6} [1-S_2(1)] \approx 0.167$, exceeding the threshold value.

We now numerically evaluate the entanglement criterion for the state (\ref{Eq:Modular_momentum_entangled_state}) when evolved under (\ref{Eq:master_equation_solution}) for the three cases (i) $x_{\Le}=-10\ell$ and $x_{\Ri}=10\ell$, (ii) $x_{\Le}=-12\ell$ and $x_{\Ri}=8\ell$, and (iii) $x_{\Le}=-13\ell$ and $x_{\Ri}=7\ell$. While all support the same initial interference pattern (\ref{Eq:Nonlocal_interference_pattern}) with $\deloc=20\ell$, (i) meets the mirror condition, whereas (ii) and (iii) exhibit increasing mismatches. In all three cases, we choose $\sigma_{x,\rel}=\ell$, $\sigma_{p,\cm}=\ell/\hbar$, and, for demonstrational purposes, strong disorder at $C_0=\hbar^2 v^2/\ell^2$. In (iv) we choose the same parameters as in (i), but with $\sigma_{x,\rel}=\ell/2$. We note that, assuming Gaussian disorder statistics, Eqs.~(\ref{Eq:master_equation_evaluated}) and (\ref{Eq:master_equation_solution}) remain valid for strong disorder \cite{Gneiting2017disorder}.

\begin{figure}[htb]
	\includegraphics[width=0.99\columnwidth]{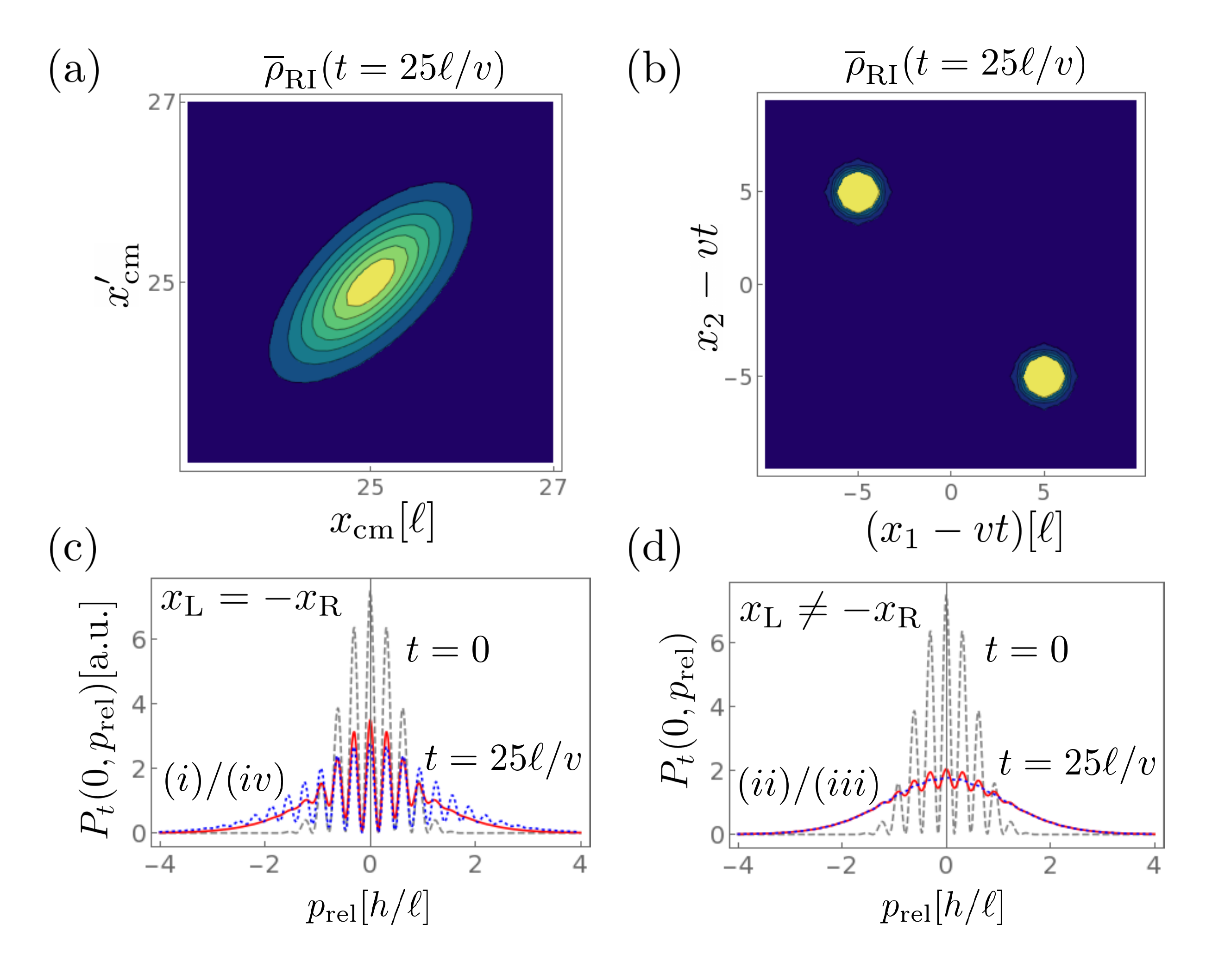}
	\caption{\label{Fig:nonlocal_interference} Disorder-perturbed evolution of the relative-state superposition (\ref{Eq:Modular_momentum_entangled_state}). (a) While the center of mass state $\bra{x_\cm,x_\Le} \overline{\rho}_{\SP}(t) \ket{x_\cm',x_\Le}$ undergoes a decay of coherences, (b) the correlations in $\bra{(x_1,x_2} \overline{\rho}_{\SP}(t) \ket{x_1,x_2}$ between the particle coordinates $x_1$ and $x_2$ remain unaffected. Depending on how well the mirror condition $x_{\Le}=-x_{\Ri}$ is met, the visibility loss in the interference pattern displayed by the relative momentum is (c) controllable or (d) exceedingly detrimental.}
\end{figure}

Figure~\ref{Fig:nonlocal_interference} shows the disorder impact at $t=25 \ell/v$, i.e., after the disorder impact has saturated. We find that, while the center-of-mass coherences decay, correlations between $N_{x,1}$ and $N_{x,2}$ remain unaffected [this follows directly from the solution (\ref{Eq:master_equation_solution})], and accordingly the corresponding variance $\langle (\Delta \hat{N}_{x, {\rm tot}})^2 \rangle$ remains close to $0$. The momentum interference, however, undergoes a mismatch-controlled visibility reduction. Notably, the interference maintains full contrast in the center of the envelope in (and only in) the mirror case. For the resulting variances $\frac{\deloc^2}{h^2} \langle (\Delta \hat{\overline{p}}_\rel)^2 \rangle$ we obtain (i) $0.136$ [red solid in (c)], (ii) $0.156$ [red solid in (d)], (iii) $0.161$ [blue dotted in (d)], and (iv) $0.120$ [blue dotted in (c)]; i.e., while in (i) and (iv) the variance remains well below, in (ii) it has reached, and in (iii) it has surpassed the entanglement detection threshold. Comparing (i) and (iv), we find that $\sigma_{x,\rel}<\ell$ further mitigates the visibility reduction, indicating that $\sigma_{x, \rel} \lessapprox \ell$ further supports disorder-robust transport, in particular in the near-dispersionless transport of edge modes.

Our previous analysis renders apparent that this robustness is independent of the delocalization $\deloc$. This confirms that the resonancelike disorder immunity at mirror points $x_\rel$ and $-x_\rel$ enables the disorder-robust transport of highly delocalized states displaying two-particle interference.

\section{$N=2$ $N00N$ state interference}

\begin{figure}[htb]
	\includegraphics[width=0.99\columnwidth]{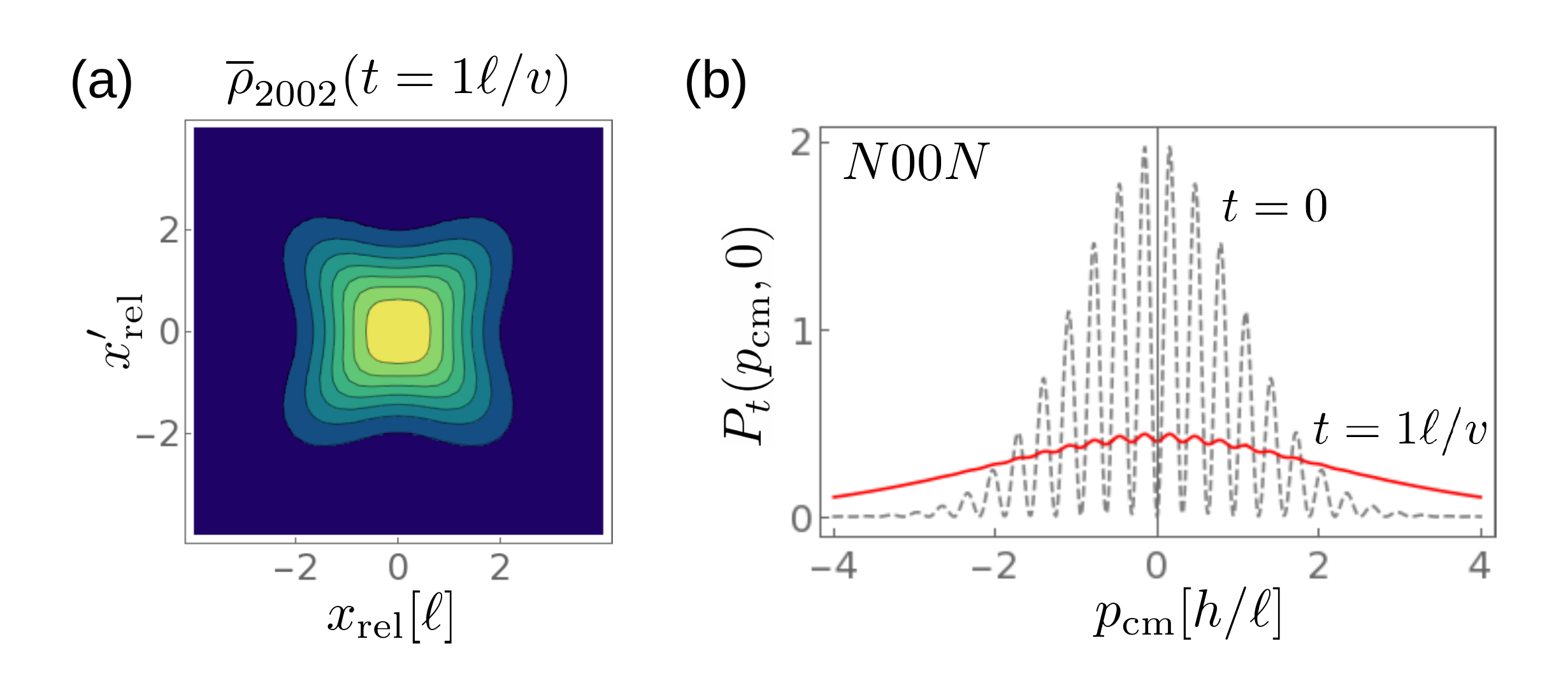}
	\caption{\label{Fig:N00N_state_interference} Disorder-perturbed transport of the two-particle ($N=2$) $N00N$ state (\ref{Eq:N00N_state}). (a) The mirror-point symmetry protects the coherences in the relative coordinate. (b) Its absence in the delocalized center-of-mass coordinate, however, causes rapid and substantial visibility loss in the center-of-mass momentum interference, highlighting the difference in the disorder sensitivity between different choices of entangled states.}
\end{figure}

To further assess the significance of the symmetry-mediated transport, we now contrast it with two-particle $N00N$ (``2002'') states, where the two particles ``bunch'' at one out of two spatially separated locations, i.e., the superposition is now in the center-of-mass coordinate:
\begin{align} \label{Eq:N00N_state}
&\ket{\Psi_{2002}} = \sqrt{\frac{\sigma_{p,\cm}}{\pi \hbar \sigma_{x,\rel}}} \int \dd x_\cm \, \dd x_\rel \, \ket{x_\cm} \otimes \ket{x_\rel} \\
&\times \frac{1}{\sqrt{2}} \left( \ee^{-\frac{\sigma_{p,\cm}^2 (x_\cm-x_{\Le})^2}{\hbar^2}} + \ee^{-\frac{\sigma_{p,\cm}^2 (x_\cm-x_{\Ri})^2}{\hbar^2}} \right) \ee^{-\frac{x_\rel^2}{4 \sigma_{x,\rel}^2}} \nonumber
\end{align}
This (symmetric) state, which can be associated with the time-bin entangled state $\frac{1}{\sqrt{2}}(\ket{\Early}_1 \ket{\Early}_2 + \ket{\Late}_1 \ket{\Late}_2)$, displays two-particle interference similar to Eq.~(\ref{Eq:Nonlocal_interference_pattern}), with the same period $\deloc$, but in the center-of-mass momentum $p_\cm$, cf.~Fig.~\ref{Fig:N00N_state_interference}.

In Fig.~\ref{Fig:N00N_state_interference}, we show the $N00N$ state (\ref{Eq:N00N_state}) when evolved under Eq.~(\ref{Eq:master_equation_solution}), with the same parameters as in case (i) above (mirror condition met). We find that, already at $t=1 \, \ell/v$ and in stark contrast to case (i) above, the visibility is strongly suppressed. Consequently, the variance of $\hat{\overline{p}}_\cm \equiv \hat{\overline{p}}_1 + \hat{\overline{p}}_2$ evaluates as $\frac{\deloc^2}{h^2} \langle (\Delta \hat{\overline{p}}_\cm)^2 \rangle \approx 0.159$, exceeding the entanglement detection threshold of the corresponding criterion $\langle (\Delta \hat{N}_{x, \rel})^2 \rangle + \frac{\deloc^2}{h^2} \langle (\Delta \hat{\overline{p}}_\cm)^2 \rangle < 2 C_{\hat{N}_x,\hat{\overline{p}}}$, where $\hat{N}_{x, \rel} \equiv \hat{N}_{x, 1} - \hat{N}_{x, 2}$. This disorder sensitivity is, of course, because the delocalization in the center-of-mass coordinate is, due to the absence of the mirror-point symmetry, not protected. This highlights a significant difference in the disorder impact between different choices of time-bin entangled states.

\section{Robust entanglement in the Haldane model}

The continuum model Eq.~\eqref{Eq:Disorder-perturbed_Hamiltonian} describes the long wavelength limit of unidirectional edge states in a variety of systems, but neglects finite size effects, such as dispersive wavepacket broadening and imperfect excitation of the topological edge states. The above analysis should therefore be seen as a baseline for dephasing of quantum states. To independently verify our results and show the persistence of robust entanglement transport in smaller, discrete systems, we simulate the propagation of two-photon states in the disordered Haldane model using the Schr\"odinger equation, described by the Hamiltonian~\cite{Haldane1988Model, Jotzu2014Haldane, Rechtsman2016topological} (See also Supplemental Material)
\begin{align}
\hat{H} = \sum_{j} (\omega_{j}^{(a)}& \hat{a}^{\dagger}_{j} \hat{a}_{j} + \omega_{j}^{(b)} \hat{b}^{\dagger}_{j} \hat{b}_{j}) + t_1 \sum_{<j,k>} (\hat{a}_{j}^{\dagger} \hat{b}_k + \hat{b}_j^{\dagger} \hat{a}_k) \nonumber \\ &+ t_2 \sum_{\ll j,k\gg} ( \hat{a}_j^{\dagger} \hat{a}_k e^{i \phi_{jk} } + \hat{b}_j^{\dagger} \hat{b}_k e^{-i \phi_{jk}} ),
\end{align}
where $\hat{a}_j^{\dagger}$ ($\hat{b}_j^{\dagger}$) creates a particle on the $a$ ($b$) sublattice in unit cell $j$, $\omega_j^{(a,b)} \in [-W/2,W/2]$ are random uncorrelated potentials, $t_1, t_2$ are nearest and next-nearest neighbor hopping strengths respectively, and flux sign $\phi_{jk} = \pm \phi$ alternates between adjacent next-nearest neighbors. We use the same parameters as in Ref.~\cite{Rechtsman2016topological}: $t_1 = 1, t_2 = 0.2, \phi = -\pi/2$, for which the gap size is $6 \sqrt{3} t_2 \sin \phi \approx 2$, and a lattice size of $N_x \times N_y = 128 \times 6$ cells, with zigzag and armchair edges. We take strong disorder $W=1.5$ (comparable to the gap size and beyond the validity of any perturbative treatment) and an ensemble of 100 disorder realizations. The disorder potential is uncorrelated, but the lattice period $a = 1$ sets a characteristic length scale for momentum broadening.

\begin{figure}
	\includegraphics[width=0.99\columnwidth]{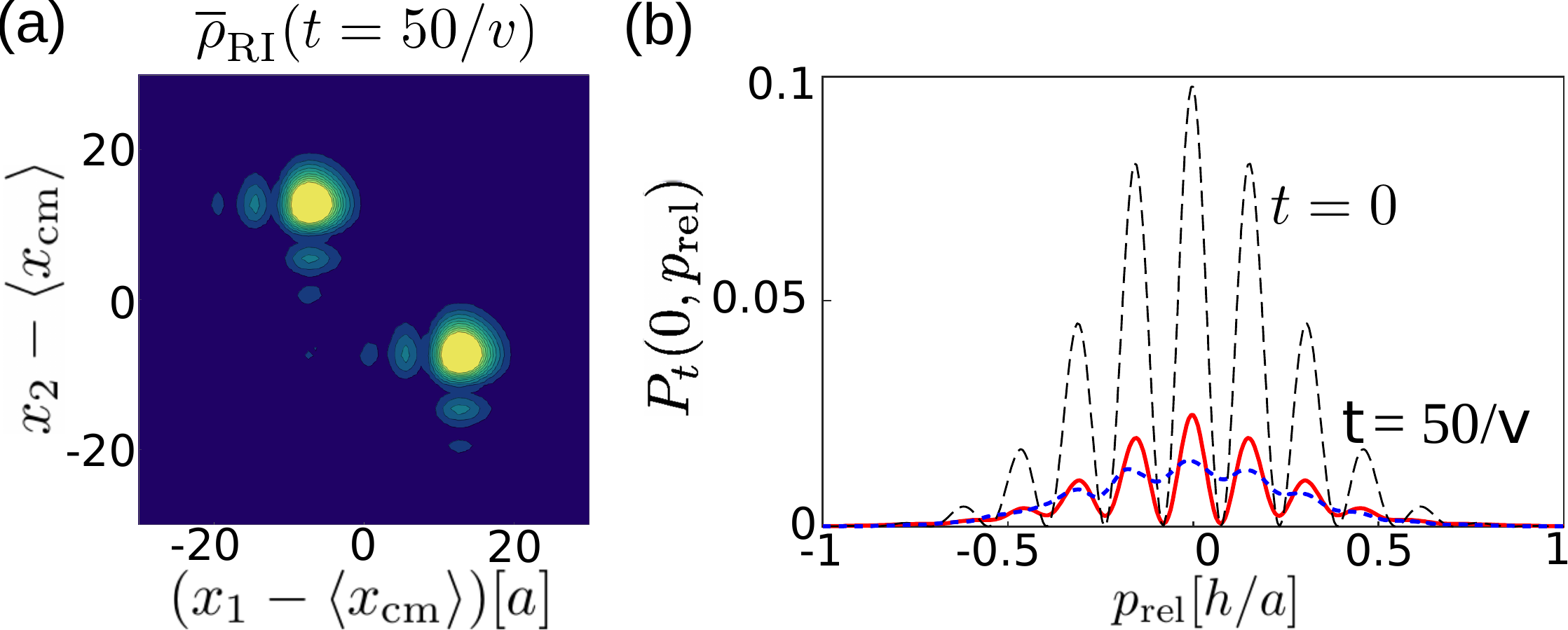}
	\caption{\label{Fig:Haldane} Disorder-perturbed evolution of relative-state superpositions Eq.~\eqref{Eq:Modular_momentum_entangled_state} in the Haldane model. (a) Correlations in real space coordinates $x_1$ and $x_2$ exhibit broadening due to the edge state dispersion. (b) Interference in the relative momentum is only robust when the mirror condition is satisfied (solid red line). The mirror-broken state $x_\Le = -32, x_\Ri = 8$ (blue dashed line) has significantly lower visibility.}
\end{figure}

We consider the initial states Eq.~\eqref{Eq:Modular_momentum_entangled_state} with $\sigma_{x,\rel}=2$, $\sigma_{p,\cm}=2$, perfect localization to the long (zigzag) edge, and tilted to excite the zigzag edge modes centred at momentum $p_{\mathrm{cm}} = \pi$ (with group velocity $v \approx 0.8$). This simple, experimentally feasible initial condition cannot perfectly excite the edge modes and some energy is lost into the bulk~\cite{Peruzzo2010quantum, Crespi2013anderson, Giuseppe2013einstein, Poulios20142D}. We compute the correlation functions along the edge after a propagation time $t=50/v$, similar to Fig.~\ref{Fig:nonlocal_interference}. The real space correlations of the mirror-symmetric state plotted in Fig.~\ref{Fig:Haldane}(a) show diffractive broadening introduced by the edge states' nonzero dispersion. Nevertheless, the mirror symmetry is preserved during propagation, resulting in robust two-particle interference in the relative momentum. Figure~\ref{Fig:Haldane}(b) reveals remarkably high visibility ($\approx 95\%$) for exact mirror symmetry ($x_\Le=-20,x_\Ri=20$) and significantly reduced visibility ($\approx 15\%$ for $x_\Le = -32, x_\Ri = 8$) for mirror-broken states. Moreover, under the same conditions $N=2$ $N00N$ states suffer an almost complete loss of interference visibility (to $\approx 8\%$) within $t=5/v$.

\section{Discussion}

We have shown, analytically and numerically, that backscattering-free disordered transport in topological edge states can exhibit a stronger form of robustness in the multiparticle case: by employing suitably chosen entangled states, one can achieve disorder-robust transport of relative phases and entanglement between spatially or temporally separated wavepackets, which is of utmost importance for applications such as interferometry and buffering of signals in quantum networks. This disorder-robust entanglement transport cannot be understood simply in terms of the familiar single particle ``immunity to backscattering'' picture. Our predictions can be readily observed by propagation of entangled two photon edge states in two-dimensional topological waveguide arrays~\cite{Kaiser2012, Matthews2013, Rechtsman2013topological, Poulios20142D, Blanco2018} or coupled resonator lattices~\cite{Mittal2016topologically, Mittal2017topologically}. Near-future electronic implementations are also conceivable \cite{Bocquillon2013coherence, Johnson2018phonon}, e.g. using spin-momentum locked quantum wires~\cite{Quay2010}.

We expect analogous conditions for disorder-robust transport hold for three or more particles, which would not only allow preservation of many-particle interference, but generally help to assess the disorder impact on multipartite interference devices [cf.~Fig~\ref{Fig:entangled_edge-mode_pair}(b)], and, ultimately, further deepen our understanding of the relation between disorder and many-particle physics beyond localization.


\paragraph{Acknowledgments.}
F.N. is partially supported by the MURI Center for Dynamic Magneto-Optics via the Air Force Office of Scientific Research (AFOSR) (FA9550-14-1-0040), the Army Research Office (ARO) (Grant No.~W911NF-18-1-0358), the Asian Office of Aerospace Research and Development (AOARD) (Grant~No.~FA2386-18-1-4045), the Japan Science and Technology Agency (JST) (the ImPACT program and CREST Grant No.~JPMJCR1676), the RIKEN-AIST Challenge Research Fund, the Japan Society for the Promotion of Science (JSPS) (JSPS-RFBR Grant No.~17-52-50023 and JSPS-FWO Grant No.~VS.059.18N), and the John Templeton Foundation. D.L. is supported by the Institute for Basic Science in Korea (IBS-R024-Y1).

\bibliography{literature}


\makeatletter 
\renewcommand{\theequation}{S\arabic{equation}}
\makeatother
\setcounter{equation}{0}

\section{Supplemental Material}

In this Supplemental Material we provide details of our numerical simulation of the Haldane model, which gives an independent validation of our analytical results. Similar to the calculations presented in the main text, we use a first quantization approach which encompasses either distinguishable or indistinguishable particles, depending on the choice of initial state. In the Haldane model, the time evolution of a single particle wavefunction $\Psi_{nm} = (a_{nm},b_{nm})$ is governed by the Schr\"odinger equation
\begin{subequations}
	\begin{align}
	i \partial_t a_{nm} &= \omega^{(a)}_{nm} a_{nm} + t_1 \sum_{\mathrm{n.n}} b_{n^{\prime} m^{\prime}} + t_2 \sum_{\mathrm{n.n.n}} a_{n^{\prime} m^{\prime}} e^{i \phi_{n^{\prime} m^{\prime}}}, \\
	i \partial_t b_{nm} &= \omega^{(b)}_{nm} b_{nm} + t_1 \sum_{\mathrm{n.n}} a_{n^{\prime} m^{\prime}} + t_2 \sum_{\mathrm{n.n.n}} b_{n^{\prime} m^{\prime}} e^{-i \phi_{n^{\prime} m^{\prime}}}, 
	\end{align}
\end{subequations}
where $(n,m)$ index the unit cells of the lattice, a honeycomb lattice formed by two sublattices $(a,b)$, $t_1$ is the nearest neighbor hopping strength, $t_2$ is the next-nearest neighbor hopping strength, signs of the fluxes $\phi_{n^{\prime} m^{\prime}} = \pm \phi$ alternate between adjacent next-nearest neighbors, and $\omega^{(j)}_{mn}$ describes uncorrelated on-site disorder uniformly distributed in the width $[-W/2,W/2]$. Fourier transforming yields the Bloch Hamiltonian,
\begin{align}
H(\bs{k}) =& 2 t_2 \cos \phi \sum_i \cos (\bs{k}\cdot \bs{a}_i) \sigma_0 \nonumber \\
&+ t_1 \sum_i [ \cos (\bs{k}\cdot \bs{\delta}_1) \sigma_x + \sin(\bs{k}\cdot \bs{\delta}_i) \sigma_y ] \\
&- 2 t_1 \sin \phi \sum_i \sin (\bs{k}\cdot \bs{a}_j)  \sigma_z, \nonumber
\end{align}
where the lattice vectors are $\bs{a}_{1,2,3}$, $\bs{\delta}_{1,2,3}$ are displacements between neighboring lattice sites, and $\sigma_j$ are Pauli matrices. The gap size is $6 \sqrt{3} t_2 \sin \phi$ with band extrema $6t_2 \cos \phi \pm 3 t_1$. 

The evolution of a two (non-interacting) particle state $\mid \Psi \rangle = \sum_{x_1,x_2} \psi_{x_1,x_2} \mid x_1 \rangle \otimes \mid x_2 \rangle$ is governed by the symmetric Hamiltonian $\hat{H}_{\mathrm{tot}} = \hat{H} \otimes \hat{1} + \hat{1}\otimes \hat{H}$. In this case, the Schr\"odinger equation reads
\begin{equation} 
i \partial_t \mid \Psi \rangle = \sum_{x_1,x_2} \psi_{x_1,x_2} ( \hat{H} \mid x_1 \rangle \otimes \mid x_2 \rangle + \mid x_1 \rangle \otimes  \hat{H} \mid x_2 \rangle ).
\end{equation}
The evolution equation for the wavefunction is obtained by multiplying both sides by $\langle r_1 \mid \otimes \langle r_2 \mid$,
\begin{align}
i \partial_t \psi_{r_1,r_2} &= \sum_{x} (\psi_{x,r_2} \langle r_1 \mid \hat{H} \mid x \rangle + \psi_{r_1,x} \langle r_2 \mid \hat{H} \mid x \rangle) \nonumber \\
&= \sum_x ( \psi_{x,r_2} H_{r_1,x} + \psi_{r_1,x} H_{r_2,x} ),
\end{align}
which is equivalent to the matrix equation
\begin{equation} 
i \partial_t \psi = H\psi + (H\psi)^T,
\end{equation}]
thus the solution is
\begin{equation}
\psi(t) = e^{-i t H } \psi(0) e^{-i t H^T}. \label{eq:evolution}
\end{equation}
We solve Eq.~\eqref{eq:evolution} for 100 different realizations of the disorder and construct the disorder-averaged density matrix as $\bar{\rho}(t) = \sum_{i=1}^{100} \mid \psi_i (t) \rangle \langle \psi_i(t) \mid$, from which we obtain the correlation functions plotted in Fig.~5 of the main text.

\end{document}